\documentclass[journal]{IEEEtran}
\IEEEoverridecommandlockouts
\usepackage[noadjust]{cite}
\usepackage{amsmath,amssymb,amsfonts}
\usepackage{algorithmic}
\usepackage{graphicx}
\usepackage{textcomp,dsfont}
\usepackage{xcolor,comment}
\usepackage{nicefrac,xfrac}
\usepackage{subcaption}
\usepackage{flushend}

\def\BibTeX{{\rm B\kern-.05em{\sc i\kern-.025em b}\kern-.08em
    T\kern-.1667em\lower.7ex\hbox{E}\kern-.125emX}}

\begin{document}

\title{Pulse Shaping Filter Design for Integrated \\ Sensing \& Communication with Zak-OTFS} 

\author{Nishant Mehrotra$^*$, Sandesh Rao Mattu$^*$, Robert Calderbank~\IEEEmembership{Life Fellow,~IEEE}
\thanks{This work is supported by the National Science Foundation under grants 2342690 and 2148212, in part by funds from federal agency and industry partners as specified in the Resilient \& Intelligent NextG Systems (RINGS) program, and in part by the Air Force Office of Scientific Research under grants FA 8750-20-2-0504 and FA 9550-23-1-0249. \\ The authors are with the Department of Electrical and Computer Engineering, Duke University, Durham, NC, 27708, USA (email: \{nishant.mehrotra,\allowbreak sandesh.mattu,\allowbreak robert.calderbank\}\allowbreak@duke.edu). \\ $*$ denotes equal contribution.}
}


\maketitle
\begin{abstract}
Zak-OTFS provides a framework for integrated sensing \& communication (ISAC) in high delay and Doppler spread environments. Pulse shaping filter design enables joint optimization of sensing and communication performance. For sensing, a localized pulse shaping filter enables input-output (I/O) relation estimates close to the physical scattering channel. For communication, orthogonality of the pulse shape on the information lattice prevents inter-symbol interference, and no time and bandwidth expansion enables full spectral efficiency. A filter simultaneously meeting all three objectives is ideal for ISAC. Existing filter designs achieve two, but not all three objectives. In this work, we design pulse shaping filters meeting all three objectives via the Isotropic Orthogonal Transform Algorithm. The proposed filters have improved \textcolor{black}{spectral efficiency, data detection and sensing performance} over existing filter choices. 
\end{abstract}


\section{Introduction}
\label{sec:intro}

\IEEEPARstart{Z}{ak}-OTFS (orthogonal time frequency space modulation)~\cite{bitspaper1,bitspaper2,otfs_book} is an emerging framework for integrated sensing and communication (ISAC) in doubly-selective environments with high delay and Doppler spreads. The Zak-OTFS carrier waveform is a pulse in the delay-Doppler (DD) domain, formally a quasi-periodic localized function termed \emph{pulsone}, whose interaction with the doubly-selective channels does not fade~\cite{bitspaper1,bitspaper2,otfs_book}. This non-fading behavior of Zak-OTFS makes it well-suited for extreme mobility environments, such as bullet trains, where legacy time-frequency modulations, such as orthogonal frequency division multiplexing (OFDM), fade. Moreover, the response to a single Zak-OTFS carrier provides an image of the scattering environment, supporting sensing in radars~\cite{bitspaper2} \textcolor{black}{and advanced transportation networks~\cite{Jamalipour2024}}.


In its original form, the DD pulsone occupies infinite time and bandwidth. For practical implementation, the DD pulsone is limited to a finite time interval $T$ and bandwidth $B$ via DD domain filtering with a \emph{pulse shaping filter}~\cite{bitspaper1,bitspaper2,otfs_book,Calderbank2025_isac,Mohammed2024_pulseshaping,Calderbank2025_interleaved,Gopalam2024_tfwindowing,Chockalingam2025_gs}. \textcolor{black}{The choice of the pulse shaping filter is key to sensing and communication performance in interference-limited environments. A localized pulse shaping filter enables accurate input-output (I/O) relation estimation (sensing). Orthogonality of the pulse shape on the information lattice prevents inter-symbol and inter-slot interference (e.g., see~\cite{Calderbank2025_uplinkrac}), and no time \& bandwidth expansion enables full spectral efficiency. A filter simultaneously meeting all three objectives is ideal for ISAC.}


Pulse shaping filters proposed in the Zak-OTFS literature~\cite{bitspaper1,bitspaper2,otfs_book,Calderbank2025_isac,Mohammed2024_pulseshaping,Calderbank2025_interleaved,Gopalam2024_tfwindowing,Chockalingam2025_gs} meet two, but not all three objectives simultaneously. For instance, the sinc filter~\cite{bitspaper1,bitspaper2,otfs_book,Calderbank2025_isac,Mohammed2024_pulseshaping} is bandwidth \& time limited and orthogonal on the information lattice, however, is not localized. Thus, it offers good data detection (communication) performance only given perfect I/O relation knowledge~\cite{bitspaper1,bitspaper2,otfs_book,Calderbank2025_isac,Mohammed2024_pulseshaping}. The \textcolor{black}{root raised cosine} (RRC) filter is localized and orthogonal, but is not bandwidth \& time limited~\cite{bitspaper1,bitspaper2,otfs_book,Calderbank2025_isac,Mohammed2024_pulseshaping,Calderbank2025_interleaved,Gopalam2024_tfwindowing}. The Gaussian filter is maximally localized and bandwidth \& time limited, but is not orthogonal~\cite{Mohammed2024_pulseshaping}, which degrades data detection performance. Recent work~\cite{Chockalingam2025_gs} utilizes the (normalized) product of the Gaussian and sinc filters, termed the Gaussian-sinc filter, which is more localized than the sinc filter and bandwidth \& time limited, but is not orthogonal on the information lattice. 

\begin{table}
    \centering
    \caption{Comparison of different DD pulse shaping filters.}
    {
    \setlength{\tabcolsep}{2.25pt}
    \renewcommand{\arraystretch}{1.25}
    \begin{tabular}{|c|c|c|c|}
         \hline
         Filter & Sidelobe Level & Orthogonal? & Time/BW Limited? \\ 
         \hline
         Sinc~\cite{bitspaper1,bitspaper2,otfs_book,Calderbank2025_isac,Mohammed2024_pulseshaping} & High & \checkmark & \checkmark \\ 
         RRC~\cite{bitspaper1,bitspaper2,otfs_book,Calderbank2025_isac,Mohammed2024_pulseshaping,Calderbank2025_interleaved,Gopalam2024_tfwindowing} & Low & \checkmark & $\times$ \\ 
         Gaussian~\cite{Mohammed2024_pulseshaping} & None & $\times$ & \checkmark \\ 
         Gaussian-sinc~\cite{Chockalingam2025_gs} & Low & $\times$ & \checkmark \\
         \textbf{IOTA (Proposed)} & \textbf{Low} & \textbf{\checkmark} & \textbf{\checkmark} \\
         \hline
    \end{tabular}
    }
    \label{tab:prior_work}
\end{table}

In this paper, we design optimal DD pulse shaping filters using the Isotropic Orthogonal Transform Algorithm (IOTA)\footnote{while possible, we do not consider the alternate approach of Hermite pulse shapes~\cite{Belfiore1997_hermite,Arslan2014_fbmcsurvey,Farhang2015_fbmc} whose performance is similar to the Gaussian-sinc filter~\cite{Chockalingam2025_hermite}}, by orthogonalizing the maximally localized\footnote{satisfying Heisenberg's Uncertainty Principle~\cite{Strohmer2003_iota,Arslan2014_fbmcsurvey,pollak1961_pswf1,pollak1961_pswf2}} Gaussian and the prolate spheroidal wave function (PSWF)~\cite{pollak1961_pswf1,pollak1961_pswf2} pulse shapes. \textcolor{black}{The IOTA procedure has been used to provide insight into characteristics of time-frequency modulations~\cite{Berrou1995_iota,Strohmer2003_iota,Arslan2014_fbmcsurvey,Farhang2015_fbmc}. Understanding the same characteristics of DD modulations is important and is novel to this paper. In Section~\ref{sec:results}, we show that the IOTA filters optimally balance the trade-off between localization and orthogonality without time \& bandwidth expansion. Our results also show that while the Gaussian-sinc filter~\cite{Chockalingam2025_gs} is near-optimal, due to its non-orthogonality, additional spectral efficiency gains of $1\%$ are possible via orthogonalization using the IOTA approach.} Table~\ref{tab:prior_work} places our contributions in the context of prior work. 


\section{Overview of Zak-OTFS}
\label{sec:prelim}

We provide a brief overview of Zak-OTFS in this Section, and refer the interested reader to~\cite{otfs_book,bitspaper1,bitspaper2} for more details.

The Zak-OTFS carrier waveform is a pulse in the delay-Doppler (DD) domain, formally a quasi-periodic localized function termed the \emph{DD pulsone}. The DD pulsone is parameterized by a delay period $\tau_p$ and a Doppler period $\nu_p$, with $\tau_p \nu_p = 1$. The DD pulsone located at delay $\tau_0$ and Doppler $\nu_0$, where $0 \leq \tau_0 < \tau_p,~0 \leq \nu_0 < \nu_p$, is given by~\cite{otfs_book,bitspaper1,bitspaper2}:
\begin{align}
    \label{eq:sys_model1}
    \mathbf{p}_{(\tau_0,\nu_0)}(\tau,\nu) &= \sum_{n,m \in \mathbb{Z}} e^{j2\pi n \nu \tau_p} \delta(\tau-n\tau_p-\tau_0) \nonumber \\ &\qquad\qquad\qquad~\times\delta(\nu-m\nu_p-\nu_0).
\end{align}

The DD pulsone in~\eqref{eq:sys_model1} occupies infinite time and bandwidth. For practical implementation, the DD pulsone is limited to a time interval $T$ and a bandwidth $B$ via DD domain filtering with a pulse shaping filter $\mathbf{w}(\tau,\nu)$~\cite{otfs_book,bitspaper1,bitspaper2}:
\begin{align}
    \label{eq:sys_model2}
    \mathbf{p}_{(\tau_0,\nu_0)}^{\mathbf{w}}(\tau,\nu) &= \mathbf{w}(\tau,\nu) *_\sigma \mathbf{p}_{(\tau_0,\nu_0)}(\tau,\nu),
\end{align}
where $*_\sigma$ denotes twisted convolution\footnote{$\mathbf{a}(\tau,\nu)*_\sigma \mathbf{b}(\tau,\nu) = \iint \mathbf{a}(\tau',\nu') \mathbf{b}(\tau-\tau',\nu-\nu') e^{j2\pi\nu'(\tau-\tau')} d\tau' d\nu'$}.

We mount $BT = MN$ information symbols on the pulse shaped pulsones in~\eqref{eq:sys_model2} at $M = \nicefrac{\tau_p}{\nicefrac{1}{B}} = B\tau_p$ distinct locations along delay and $N = \nicefrac{\nu_p}{\nicefrac{1}{T}} = T\nu_p$ distinct locations along Doppler. Specifically, information symbols are mounted on the lattice $\Lambda = \big\{(\tau_0,\nu_0) = \big(\nicefrac{k_0}{B},\nicefrac{l_0}{T}\big): k_0 \in \mathbb{Z}_{M},l_0 \in \mathbb{Z}_{N}\big\}$ as:
\begin{align}
    \label{eq:sys_model3}
    \mathbf{x}^{\mathbf{w}}_{{}_\mathsf{DD}}(\tau,\nu) &= \sum_{k_0=0}^{M-1} \sum_{l_0=0}^{N-1} \mathbf{X}[k_0,l_0] \mathbf{p}_{(\nicefrac{k_0}{B},\nicefrac{l_0}{T})}^{\mathbf{w}}(\tau,\nu) \nonumber \\
    &= \mathbf{w}(\tau,\nu) *_\sigma \mathbf{x}_{{}_\mathsf{DD}}(\tau,\nu),
\end{align}
where $\mathbf{X}$ is the $M \times N$ array of information symbols, and:
\begin{align}
    \label{eq:sys_model5}
    \mathbf{x}_{{}_\mathsf{DD}}(\tau,\nu) &= \sum_{k_0=0}^{M-1} \sum_{l_0=0}^{N-1} \mathbf{X}[k_0,l_0] \mathbf{p}_{(\nicefrac{k_0}{B},\nicefrac{l_0}{T})}(\tau,\nu).
\end{align}


After pulse shaping, we transmit the time domain (TD) representation of the signal $\mathbf{x}^{\mathbf{w}}_{{}_\mathsf{DD}}(\tau,\nu)$, generated via the inverse Zak transform\footnote{$\mathbf{a}(t) = \mathcal{Z}^{-1}(\mathbf{a}(\tau,\nu)) = \sqrt{\tau_p} \int_{0}^{\nu_p} \mathbf{a}(t,\nu) d\nu$}. Let $\mathbf{h}_{\mathrm{phy}}(\tau,\nu) = \sum_{i=1}^{P} h_i \delta(\tau-\tau_i) \delta(\nu-\nu_i)$ denote the DD representation of a scattering environment with $P$ paths. The receiver performs matched filtering with the filter $\mathbf{\tilde{w}}(\tau,\nu) = e^{j2\pi\nu\tau} \mathbf{w}^*(-\tau,-\nu)$ to obtain~\cite{otfs_book,bitspaper1,bitspaper2}:
\begin{align}
    \label{eq:sys_model7}
    \mathbf{y}^{\mathbf{\tilde{w}}}_{{}_\mathsf{DD}}(\tau,\nu) &= \mathbf{h}_{\mathrm{eff}}(\tau,\nu) *_\sigma \mathbf{x}_{{}_\mathsf{DD}}(\tau,\nu) + \mathbf{n}^{\mathbf{\tilde{w}}}_{{}_\mathsf{DD}}(\tau,\nu), 
    &= 
\end{align}
where $\mathbf{h}_{\mathsf{eff}}(\tau,\nu)$ denotes the \emph{effective channel}\footnote{the effective channel approximates the physical channel when all paths are resolvable in delay with bandwidth $B$ and in Doppler with time $T$} that encompasses the effects of the physical scattering environment as well as pulse shaping and matched filtering~\cite{otfs_book,bitspaper1,bitspaper2}:
\begin{align}
    \label{eq:sys_model8}
    \mathbf{h}_{\mathrm{eff}}(\tau,\nu) &= \mathbf{\tilde{w}}(\tau,\nu) *_\sigma \mathbf{h}_{\mathrm{phy}}(\tau,\nu) *_\sigma \mathbf{w}(\tau,\nu),
\end{align}
and $\mathbf{n}^{\mathbf{\tilde{w}}}_{{}_\mathsf{DD}}(\tau,\nu)$ denotes receiver noise after matched filtering. 

Upon sampling on the information symbol lattice $\Lambda$, the system model in~\eqref{eq:sys_model7} reduces to~\cite{otfs_book,bitspaper1,bitspaper2}:
\begin{align}
    \label{eq:sys_model9}
    \mathbf{y}^{\mathbf{\tilde{w}}}_{{}_\mathsf{DD}}[k,l] &= \mathbf{h}_{\mathrm{eff}}[k,l] *_{\sigma_{{}_{d}}} \mathbf{x}_{{}_\mathsf{DD}}[k,l] + \mathbf{n}^{\mathbf{\tilde{w}}}_{{}_\mathsf{DD}}[k,l],
\end{align}
where $\mathbf{a}[k,l] = \mathbf{a}\big(\nicefrac{k}{B},\nicefrac{l}{T}\big),k \in \mathbb{Z}_{M},l \in \mathbb{Z}_{N}$, denotes samples of $\mathbf{a}(\tau,\nu) \in \big\{\mathbf{y}^{\mathbf{\tilde{w}}}_{{}_\mathsf{DD}}(\tau,\nu), \mathbf{h}_{\mathrm{eff}}(\tau,\nu), \mathbf{x}_{{}_\mathsf{DD}}(\tau,\nu), \mathbf{n}^{\mathbf{\tilde{w}}}_{{}_\mathsf{DD}}(\tau,\nu) \big\}$ on the lattice $\Lambda$, and $*_{\sigma_{{}_{d}}}$ denotes discrete twisted convolution\footnote{$\mathbf{a}[k,l]*_{\sigma_{{}_{d}}} \mathbf{b}[k,l] = \sum_{k',l'\in\mathbb{Z}} \mathbf{a}[k-k',l-l'] \mathbf{b}[k',l'] e^{\frac{j2\pi}{MN}k'(l-l')}$}.

Substituting~\eqref{eq:sys_model5} in~\eqref{eq:sys_model9} and vectorizing gives~\cite{otfs_book,bitspaper1,bitspaper2}:
\begin{align}
    \label{eq:sys_model10}
    \mathbf{y} &= \mathbf{H} \mathbf{x} + \mathbf{n},
\end{align}
where $\mathbf{y},\mathbf{x},\mathbf{n}$ are $MN \times 1$ vectors with $\mathbf{y}[k+lM] = \mathbf{y}^{\mathbf{\tilde{w}}}_{{}_\mathsf{DD}}[k,l]$, $\mathbf{x}[k+lM] = \mathbf{x}_{{}_\mathsf{DD}}[k,l]$, $\mathbf{n}[k+lM] = \mathbf{n}^{\mathbf{\tilde{w}}}_{{}_\mathsf{DD}}[k,l]$, and $\mathbf{H}$ is the $MN \times MN$ I/O relation matrix with entries:
\begin{align}
    \label{eq:sys_model11}
    \mathbf{H}[k\!+\!lM,\!k'\!+\!l'M]\!&=\!\sum_{n,m \in \mathbb{Z}} e^{\frac{j2\pi}{N}nl'} e^{\frac{j2\pi}{MN}(k'+nM)(l-l'-mN)} \nonumber \\ &\times \mathbf{h}_{\mathrm{eff}}[k-k'-nM,l-l'-mN].
\end{align}

Assuming the I/O relation matrix $\mathbf{H}$ is known, the receiver detects the information symbols $\mathbf{x}$ from $\mathbf{y}$, e.g., via the minimum mean squared error (MMSE) estimator~\cite{otfs_book}. To acquire the I/O relation, a known pilot symbol $\mathbf{x}_{\mathsf{p}}[k,l]$ is transmitted in one of three ways: in a frame separate from data~\cite{bitspaper2}, embedded in the data frame with appropriate guard regions~\cite{Mohammed2024_pulseshaping,Calderbank2025_interleaved}, or overlayed on the data frame via a mutually unbiased filter~\cite{Calderbank2025_isac}. \textcolor{black}{The receiver estimates the effective channel $\mathbf{h}_{\mathrm{eff}}[k,l]$ via the cross-ambiguity function\footnote{\textcolor{black}{Cross-ambiguity-based effective channel estimates are equivalent to range-Doppler maps in radar sensing. For more details, see~\cite{EURASIP2025,otfs_book,bitspaper2,Calderbank2025_isac,zakotfs_ltv}.}} between the received and transmitted pilot symbols~\cite{Calderbank2025_isac}:}
\begin{align}
    \label{eq:sys_model12}
    \textcolor{black}{\widehat{\mathbf{h}}_{\mathrm{eff}}[k,l]} &\textcolor{black}{=} \textcolor{black}{\frac{1}{MN} \sum_{k'=0}^{M-1}\sum_{l'=0}^{N-1}\mathbf{y}^{\mathbf{\tilde{w}}}_{{}_\mathsf{DD}}[k', l']\mathbf{x}_{\mathsf{p}}^*[k'-k, l'-l]} \nonumber \\ &\qquad\qquad\qquad\quad \textcolor{black}{\times e^{-\frac{j2\pi}{MN}l(k'-k)},}
\end{align}
which is subsequently used to estimate the matrix $\mathbf{H}$ via~\eqref{eq:sys_model11}.

\section{Pulse Shaping Filters \& Their Design}
\label{sec:filter_design}

\begin{figure*}
    \centering
    \begin{subfigure}{0.47\linewidth}
    \includegraphics[width=\textwidth]{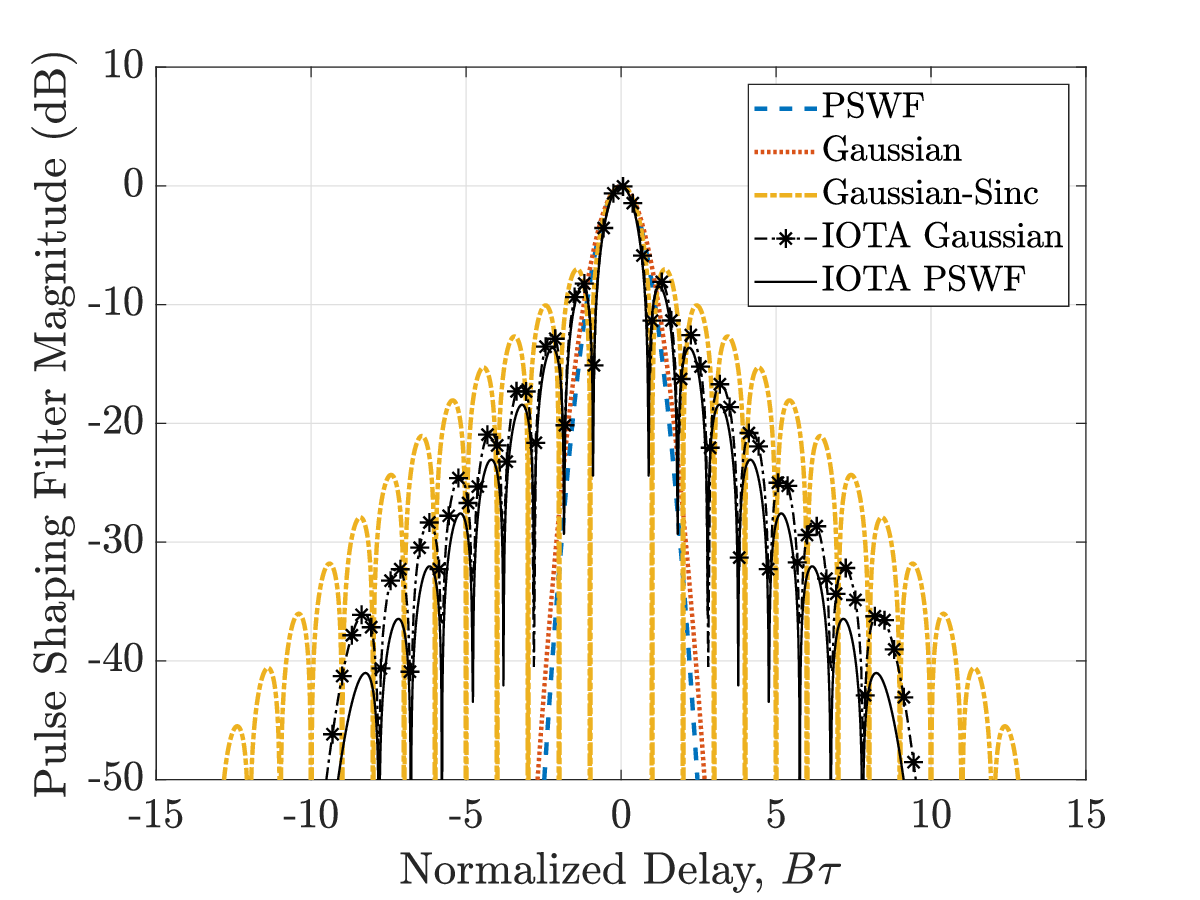}
    \caption{Pulse shaping filter magnitude vs normalized delay.}
        \label{fig:pulse_vs_del}
    \end{subfigure}
    \begin{subfigure}{0.47\linewidth}
        \includegraphics[width=\textwidth]{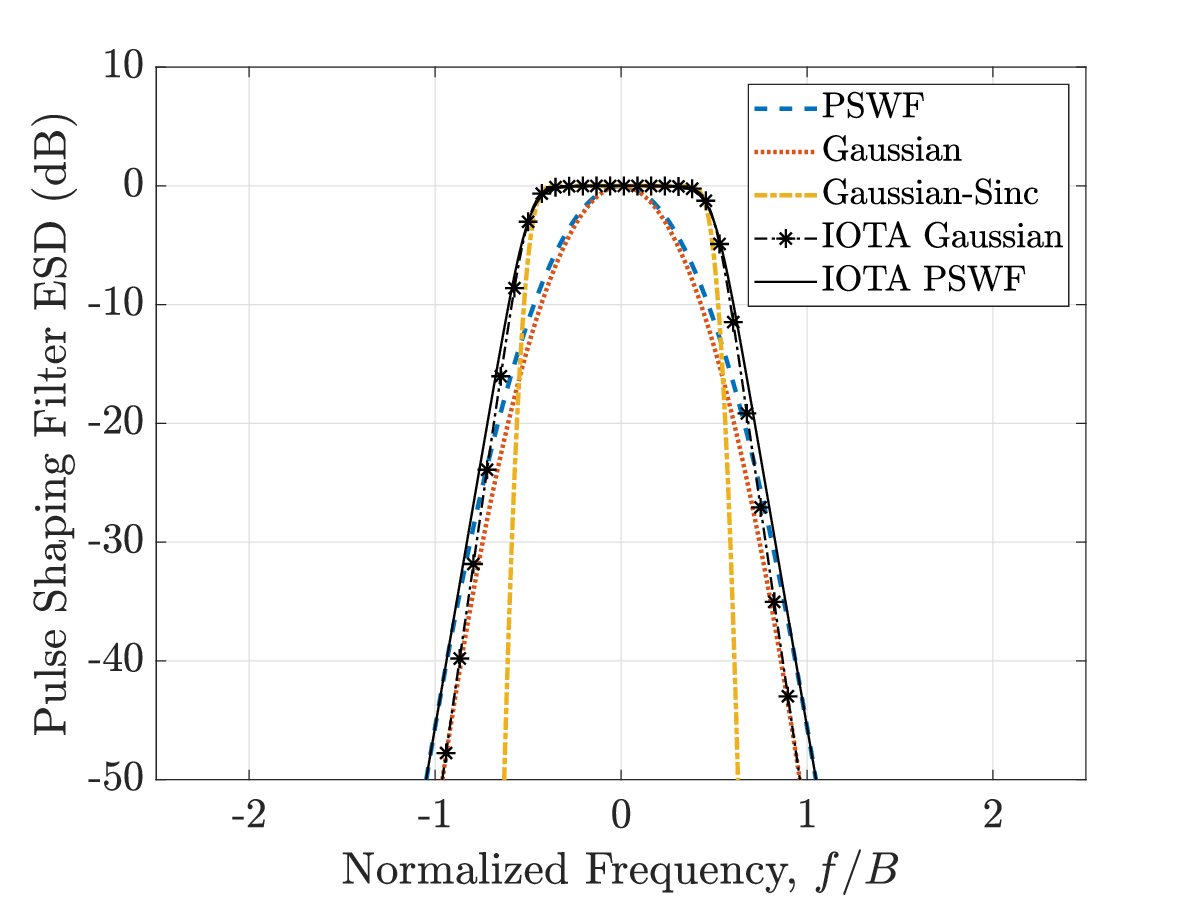}
    \caption{Energy spectral density (ESD) vs normalized frequency.}
        \label{fig:fft_vs_freq}
    \end{subfigure}

    \vspace{0.5em}

    \begin{subfigure}{0.47\linewidth}
    \includegraphics[width=\textwidth]{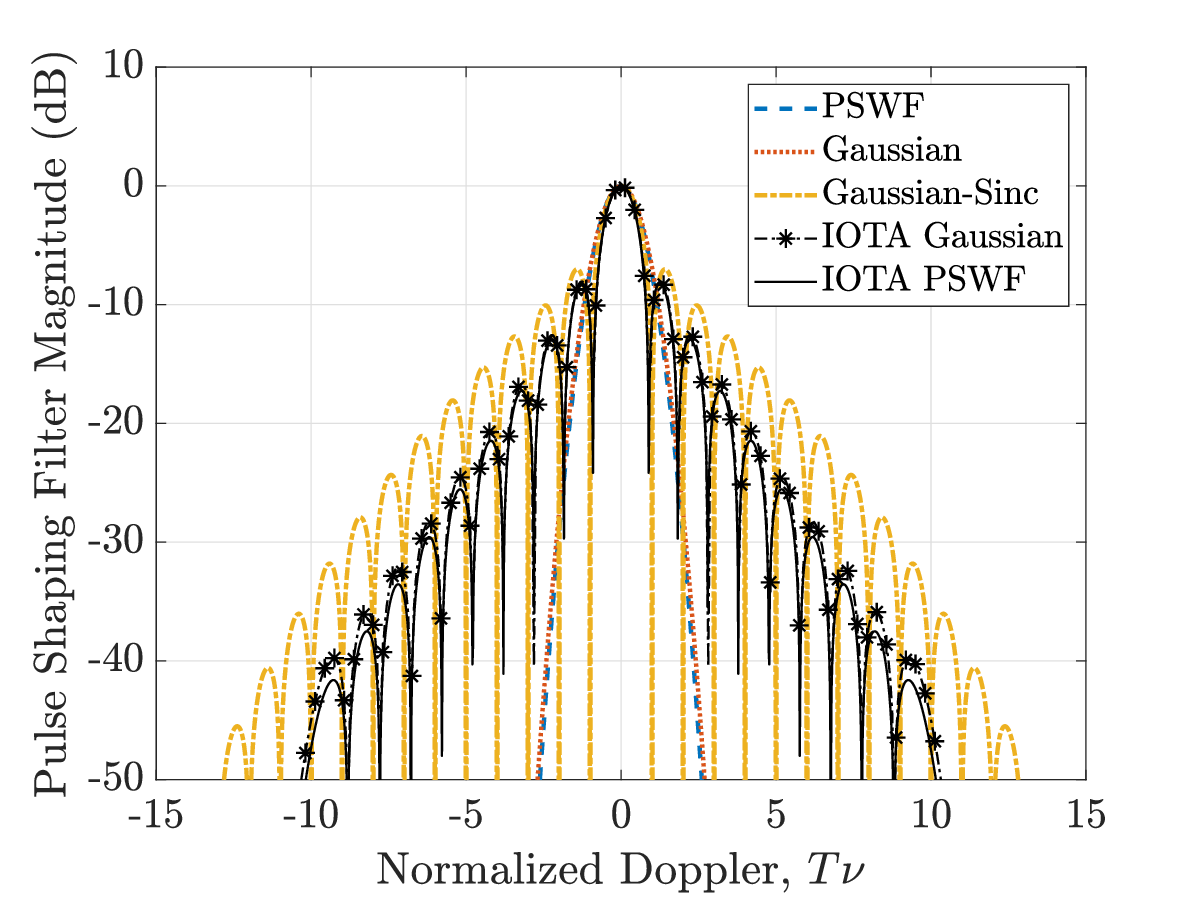}
    \caption{Pulse shaping filter magnitude vs normalized Doppler.}
        \label{fig:pulse_vs_dop}
    \end{subfigure}
    \begin{subfigure}{0.47\linewidth}
        \includegraphics[width=\textwidth]{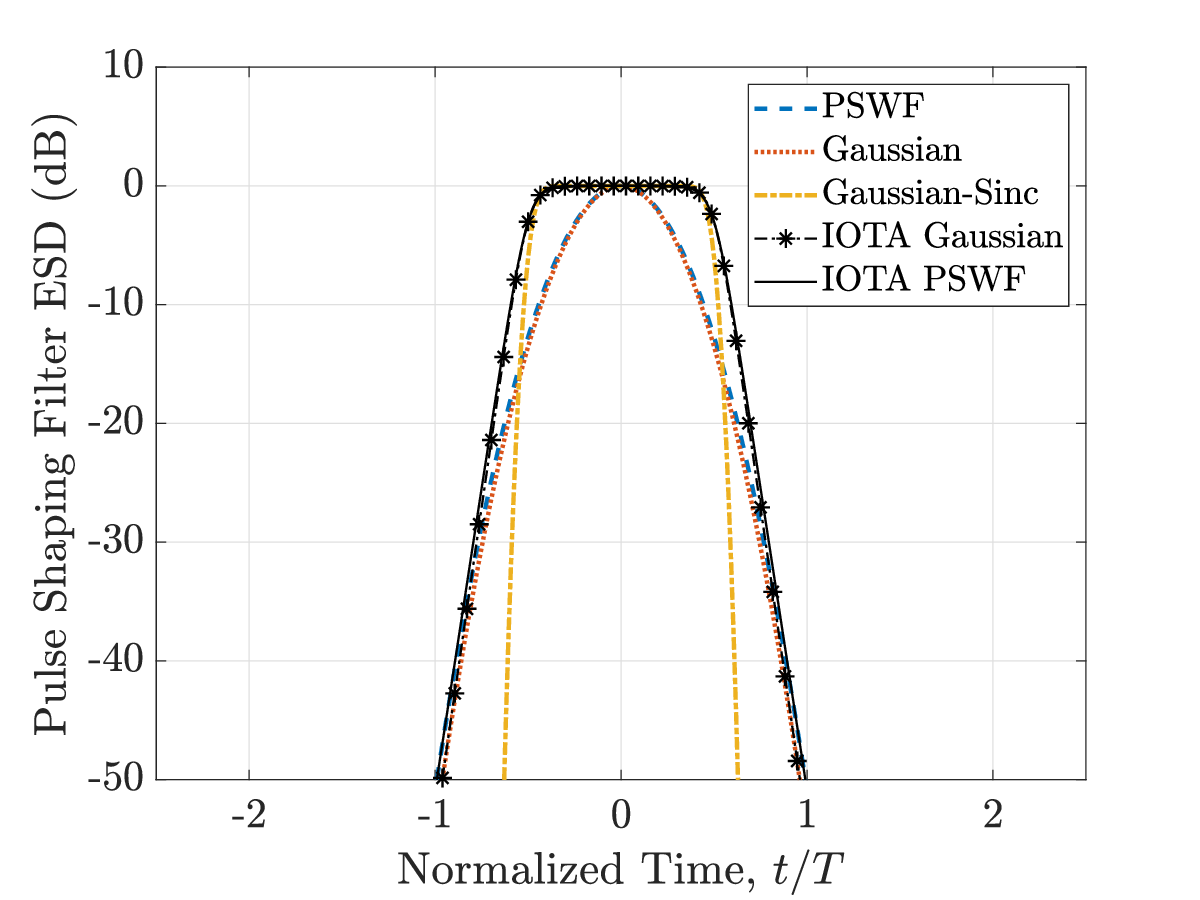}
    \caption{Energy spectral density (ESD) vs normalized time.}
        \label{fig:fft_vs_time}
    \end{subfigure}    
    
    \caption{Pulse shaping filter variation and energy concentration. (a)-(b): The proposed IOTA pulse shapes have a sharper main lobe and lower sidelobes compared to the Gaussian-sinc filter~\cite{Chockalingam2025_gs}. Moreover, the IOTA PSWF pulse has smaller main lobe and sidelobes compared to the IOTA Gaussian filter. (c)-(d): The IOTA pulse shapes have flat energy spectral density within $f \in [\nicefrac{-B}{2},\nicefrac{B}{2}]$ and $t \in [\nicefrac{-T}{2},\nicefrac{T}{2}]$ and a roll-off similar to their prototype pulses (Gaussian or PSWF). All pulse shapes have $99.99\%$ energy concentration within $f \in [\nicefrac{-B}{2},\nicefrac{B}{2}]$ and $t \in [\nicefrac{-T}{2},\nicefrac{T}{2}]$.}
    \label{fig:pulse}
\end{figure*}

In this Section, we first overview existing choices for the DD pulse shaping filter $\mathbf{w}(\tau,\nu)$ in~\eqref{eq:sys_model2}, before outlining our proposed framework for DD pulse shaping filter design.

\subsection{Standard Pulse Shaping Filters}
\label{subsec:filter_prev}

The following pulse shaping filters have been considered in prior work on Zak-OTFS~\cite{bitspaper1,bitspaper2,otfs_book,Calderbank2025_isac,Mohammed2024_pulseshaping,Calderbank2025_interleaved,Gopalam2024_tfwindowing,Chockalingam2025_gs}.

\subsubsection{Sinc~\cite{bitspaper1,bitspaper2,otfs_book,Calderbank2025_isac,Mohammed2024_pulseshaping}}

The sinc filter is given by:
\begin{align}
    \label{eq:sinc1}
    \mathbf{w}(\tau,\nu) &= \sqrt{BT}~\text{sinc}(B\tau)~\text{sinc}(T\nu).
\end{align}

\subsubsection{RRC~\cite{bitspaper1,bitspaper2,otfs_book,Calderbank2025_isac,Mohammed2024_pulseshaping,Calderbank2025_interleaved,Gopalam2024_tfwindowing}}

The RRC filter is given by:
\begin{align}
    \label{eq:rrc1}
    \mathbf{w}(\tau,\nu) &= \sqrt{BT}~\text{rrc}_{\beta_\tau}(B\tau)~\text{rrc}_{\beta_\nu}(T\nu),
\end{align}
where $0 \leq \beta_\tau,\beta_\nu \leq 1$ and:
\begin{align}
    \label{eq:rrc2}
    \text{rrc}_{\beta}(x) &= \frac{\sin(\pi x (1-\beta)) + 4\beta x\cos(\pi x (1+\beta))}{\pi x (1-(4\beta x)^2)}.
\end{align}

When $\beta_\tau = \beta_\nu = 0$, the RRC filter specializes to the sinc filter. However, when $\beta_\tau, \beta_\nu > 0$, there is bandwidth and time expansion beyond $B$ and $T$ to $(1+\beta_\tau) B$ and $(1+\beta_\nu) T$.

\subsubsection{Gaussian~\cite{Mohammed2024_pulseshaping}}

The Gaussian filter is given by:
\begin{align}
    \label{eq:gauss1}
    \mathbf{w}(\tau,\nu) &= \sqrt{BT}\bigg(\frac{4\alpha_\tau\alpha_\nu}{\pi^2}\bigg)^{\nicefrac{1}{4}} e^{-\big[\alpha_\tau(B\tau)^2+\alpha_\nu(T\nu)^2\big]}.
\end{align}

When $\alpha_\tau = \alpha_\nu = 1.584$, there is no bandwidth and time expansion beyond $B$ and $T$.

\subsubsection{Gaussian-sinc~\cite{Chockalingam2025_gs}}

The Gaussian-sinc filter is given by:
\begin{align}
    \label{eq:gs1}
    \mathbf{w}(\tau,\nu) &= \sqrt{BT}~\Omega_\tau\Omega_\nu~\text{sinc}(B\tau)~\text{sinc}(T\nu) \nonumber \\ &~~~\times e^{-\big[\alpha_\tau(B\tau)^2+\alpha_\nu(T\nu)^2\big]}.
\end{align}

When $\alpha_\tau = \alpha_\nu = 0.044$ and $\Omega_\tau = \Omega_\nu = 1.0278$, there is no bandwidth and time expansion beyond $B$ and $T$.

\subsection{IOTA Pulse Shaping Design Framework}
\label{subsec:filter_prop}

To address the limitations of existing pulse shaping filters from Section~\ref{subsec:filter_prev} (see Section~\ref{sec:intro} and Table~\ref{tab:prior_work}), we utilize the Isotropic Orthogonal Transform Algorithm (IOTA)~\cite{Berrou1995_iota,Strohmer2003_iota,Farhang2015_fbmc,Arslan2014_fbmcsurvey} framework to design DD pulse shaping filters. The key idea is to orthogonalize a maximally localized pulse shape to shifts of the lattice points $\Lambda$. The resulting pulse constitutes the closest orthogonal pulse shape to the initially chosen pulse.

To apply the IOTA procedure, we consider two maximally localized pulses -- the Gaussian pulse in~\eqref{eq:gauss1} and the prolate spheroidal wave function (PSWF)~\cite{pollak1961_pswf1,pollak1961_pswf2}. In 1D, PSWFs are functions that are bandlimited to $\big[\nicefrac{-B'}{2},\nicefrac{B'}{2}\big]$ and are maximally localized within the time interval $\big[\nicefrac{-T'}{2},\nicefrac{T'}{2}\big]$. 1D PSWFs are given by the eigenvectors of the Fredholm integral:
\begin{align}
    \label{eq:pswf1}
    \int_{\nicefrac{-T'}{2}}^{\nicefrac{T'}{2}} \text{sinc}(B'(t-t')) \psi_{n}(t') dt' &= \lambda_{n} \psi_{n}(t),
\end{align}
where $t \in \big[\nicefrac{-T'}{2},\nicefrac{T'}{2}\big]$ and $\lambda_{n}$ denotes the eigenvalue. To generate 2D PSWFs in the DD domain, we solve the problem in~\eqref{eq:pswf1} over delay and Doppler separately. Specifically, along delay, we set $B' = B$, $T' = \tau_p$, $t = \tau$, $t' = \tau'$ to obtain $\psi_{n}^{\mathsf{del}}(\tau)$. Along Doppler, we set $B' = T$, $T' = \nu_p$, $t = \nu$, $t' = \nu'$ to obtain $\psi_{n}^{\mathsf{Dop}}(\nu)$. The DD PSWF filter is given by the product $\mathbf{w}(\tau,\nu) = \psi_{n}^{\mathsf{del}}(\tau) \psi_{n}^{\mathsf{Dop}}(\nu)$, with $n = 1$ in practice.

Subsequently, we apply the IOTA procedure in two stages.

\noindent \textit{Stage 1:} First, the prototype filter (Gaussian or PSWF) is mounted on the information lattice $\Lambda$. For a prototype filter $\mathbf{w}_{\mathsf{p}}(\tau,\nu)$, the following $MN \times |\mathcal{T}||\mathcal{V}|$ matrix is formed:
\begin{align}
    \label{eq:iota1}
    \mathbf{G} = \begin{bmatrix}
        \cdots & \mathbf{w}_{\mathsf{p}}\big(\tau-\nicefrac{0}{B},\nu-\nicefrac{0}{T}\big) & \cdots \\
        & \vdots & \\
        \cdots & \mathbf{w}_{\mathsf{p}}\big(\tau-\nicefrac{k}{B},\nu-\nicefrac{l}{T}\big) & \cdots \\
        & \vdots & \\
        \cdots & \mathbf{w}_{\mathsf{p}}\big(\tau-\nicefrac{(M-1)}{B},\nu-\nicefrac{(N-1)}{T}\big) & \cdots
    \end{bmatrix},
\end{align}
where $\mathcal{T}$ denotes the set of delay values and $\mathcal{V}$ denotes the set of Doppler values spanned by the prototype filter $\mathbf{w}_{\mathsf{p}}(\tau,\nu)$.

\noindent \textit{Stage 2:} Next, the inter-pulse interference between the $MN$ rows of the matrix $\mathbf{G}$ in~\eqref{eq:iota1} is removed, such that the resulting pulses are orthogonal. Let $\mathbf{R} = \mathbf{G} \mathbf{G}^{\mathsf{H}}$ denote the Gram matrix corresponding to $\mathbf{G}$. The orthogonalized pulses are given by:
\begin{align}
    \label{eq:iota2}
    \tilde{\mathbf{G}} &= \mathbf{R}^{\nicefrac{-1}{2}} \mathbf{G},
\end{align}
\textcolor{black}{with no inter-pulse interference on the lattice $\Lambda$} since:
\begin{align}
    \label{eq:iota3}
    \tilde{\mathbf{R}} &= \tilde{\mathbf{G}} \tilde{\mathbf{G}}^{\mathsf{H}} = \mathbf{R}^{\nicefrac{-1}{2}} \mathbf{G} \mathbf{G}^{\mathsf{H}} \mathbf{R}^{\nicefrac{-1}{2}} = \mathbf{I}_{MN}.
\end{align}

The DD IOTA pulse shaping filter $\mathbf{w}(\tau,\nu)$ corresponds to the first row of $\tilde{\mathbf{G}}$ centered at zero delay and zero Doppler.

\section{Numerical Results}
\label{sec:results}

\begin{table}[!t]
    \centering
    \caption{Power-delay profile of Veh-A channel model}
    \begin{tabular}{|c|c|c|c|c|c|c|}
         \hline
         Path index $i$ & $1$ & $2$ & $3$ & $4$ & $5$ & $6$ \\
         \hline
         Delay $\tau_i (\mu s)$ & $0$ & $0.31$ & $0.71$ & $1.09$ & $1.73$ & $2.51$ \\
         \hline
         Relative power (dB) & $0$ & $-1$ & $-9$ & $-10$ & $-15$ & $-20$ \\
         \hline
    \end{tabular}
    \label{tab:veh_a}
\end{table}

\subsection{Simulation Configuration}
\label{subsec:sim_config}

We conduct numerical simulations using a 3GPP-compliant $P=6$ path Vehicular-A (Veh-A) channel model~\cite{veh_a} with significant mobility and delay spread. \textcolor{black}{The channel gain of each path is sampled from the power-delay profile in Table~\ref{tab:veh_a} with random phases}, and the Doppler of each path is simulated as $\nu_i = \nu_{\max}\cos(\theta_i)$, with $\theta_i$ uniformly distributed in $[-\pi, \pi)$ and $\nu_{\max}$ denoting the maximum channel Doppler spread\footnote{Note that our channel model is representative of real propagation environments with \textit{fractional} delay and Doppler shifts -- the path delays in Table~\ref{tab:veh_a} are non-integer multiples of the delay resolution $\nicefrac{1}{B}$, and the Doppler shifts $\nu_i = \nu_{\max}\cos(\theta_i)$ are non-integer multiples of the Doppler resolution $\nicefrac{1}{T}$.}. We simulate using the parameters: $M=17, N=19$, $\nu_{\max} = 815$ Hz\footnote{\textcolor{black}{at a carrier frequency of $3$ GHz, this corresponds to a highly dynamic scenario with maximum speed of $81.5$ m/s, e.g., a high-speed train}}, $\nu_p = 30$ kHz and $\beta_\tau = \beta_\nu = 0.6$ for the RRC filter. We do not pursue simulations with larger frame sizes and wider bandwidths here, since we have shown there is no difference in performance in~\cite{Mehrotra2025_WCLSpread}.

\begin{figure}
    \centering
    \includegraphics[width=0.95\linewidth]{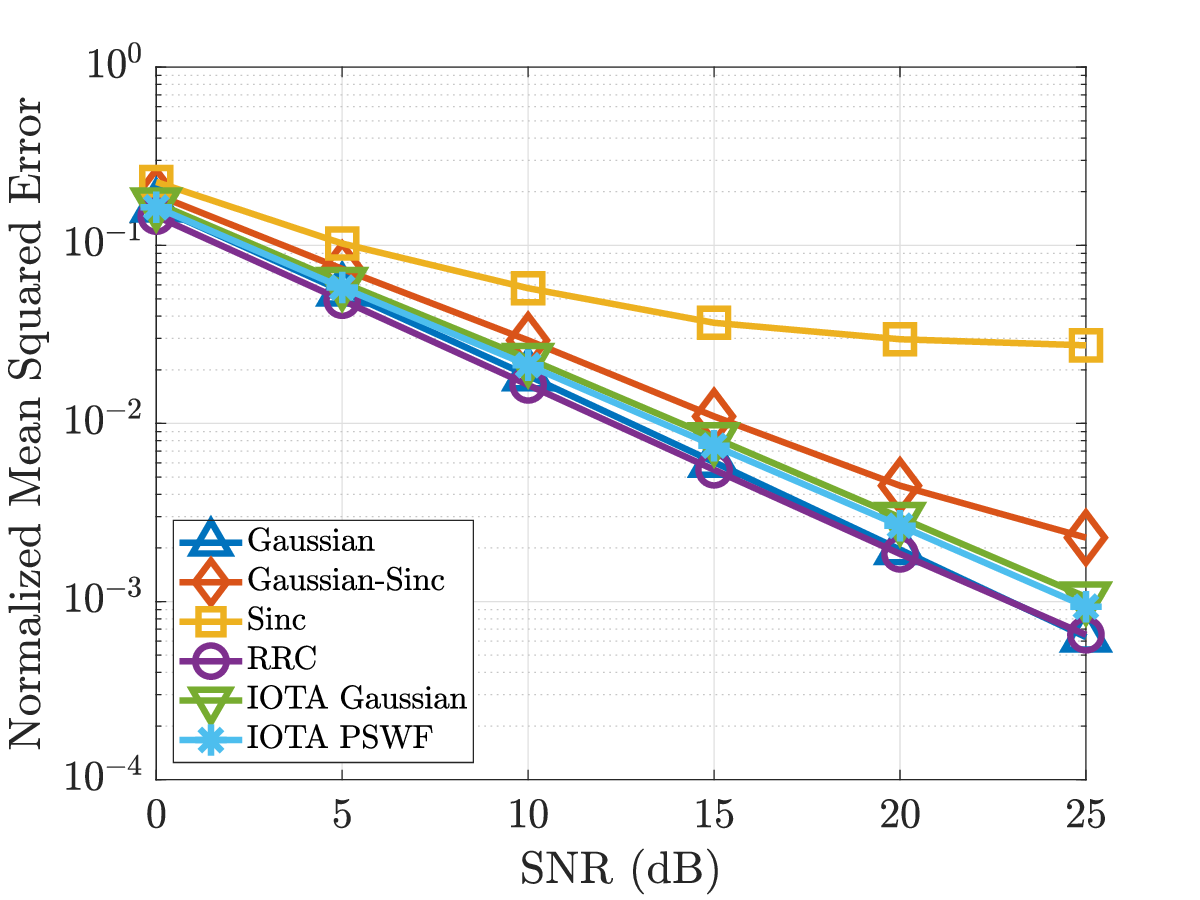}
    \caption{\textcolor{black}{Normalized mean squared error (NMSE) for estimating the DD effective channel $\mathbf{h}_{\mathrm{eff}}[k,l]$ (sensing) via~\eqref{eq:sys_model12}.}}
    \label{fig:nmse}
\end{figure}

\begin{figure*}
    \centering
    \begin{subfigure}{0.47\linewidth}
    \includegraphics[width=\textwidth]{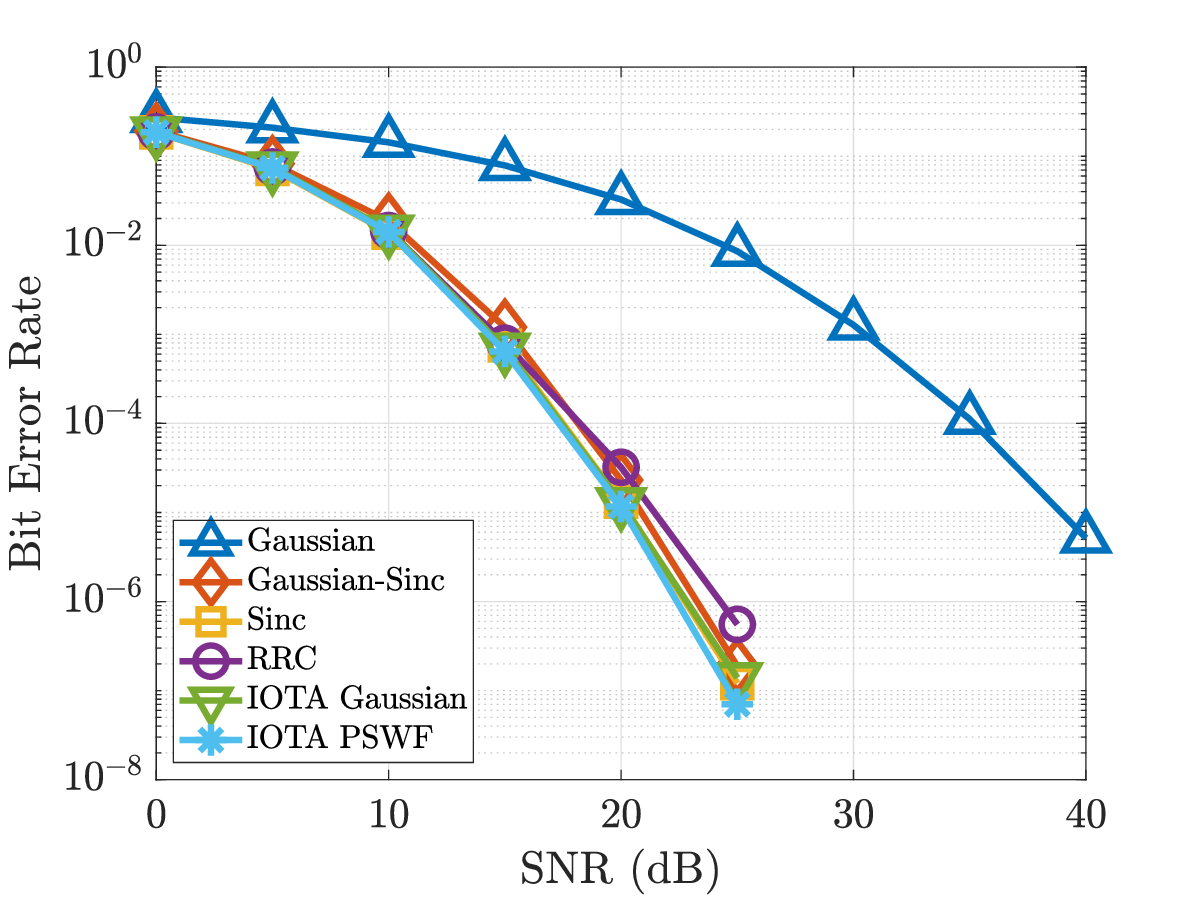}
    \caption{Bit error rate (BER) with perfect I/O relation knowledge.}
        \label{fig:ber_perfectcsi}
    \end{subfigure}
    \begin{subfigure}{0.47\linewidth}
    \includegraphics[width=\textwidth]{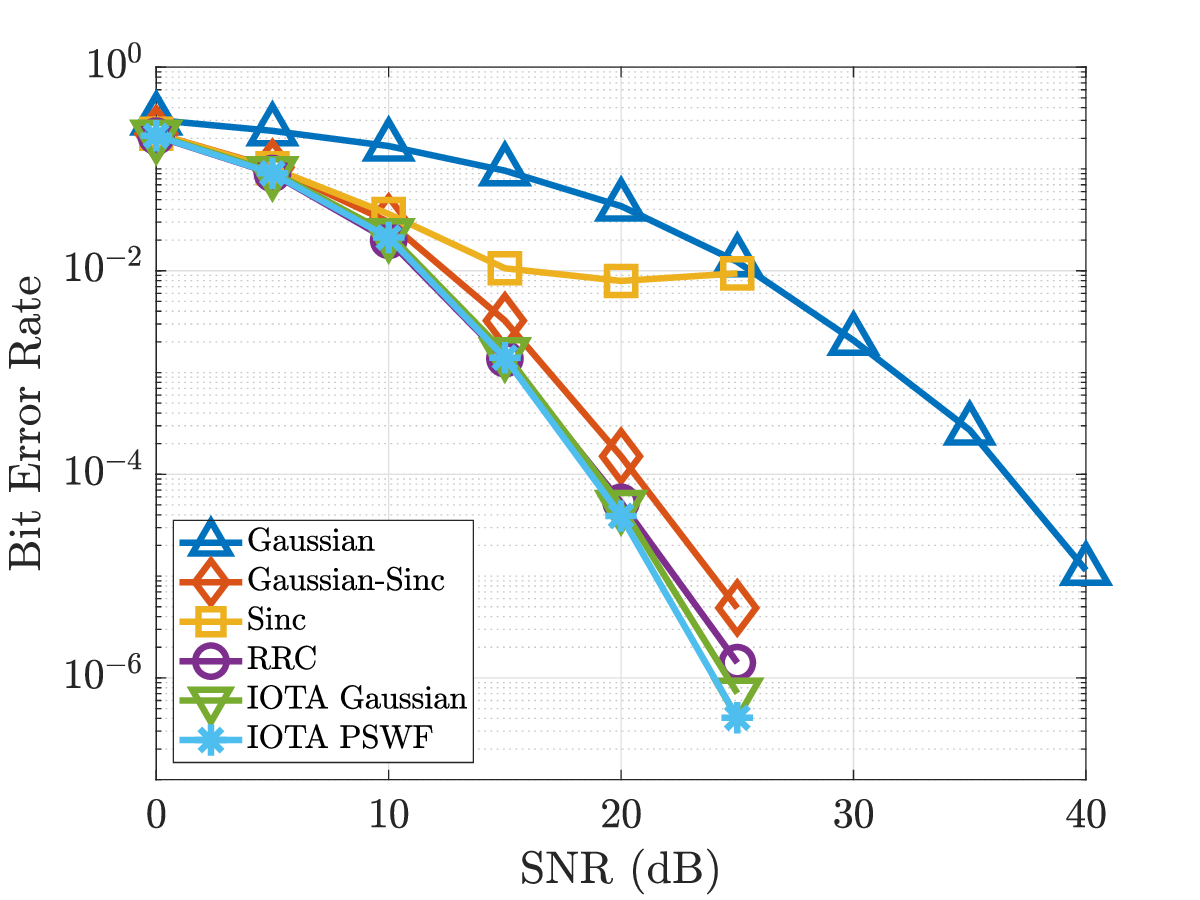}
    \caption{BER with estimated I/O relation.}
        \label{fig:ber_estcsi}
    \end{subfigure}
    \caption{\textcolor{black}{Uncoded $4$-QAM data detection performance with perfect and estimated I/O relation. The simultaneous localization, orthogonality and improved I/O relation estimation with the proposed IOTA pulses results in near-ideal BER performance.}}
    \label{fig:ber}
\end{figure*}

\subsection{Pulse Shaping Filter Visualization}
\label{subsec:pulse}

Fig.~\ref{fig:pulse} visualizes the pulse shaping filter magnitudes and their energy spectral densities (ESD), excluding the RRC pulse. All pulse shapes have $99.99\%$ energy concentration within $f \in [\nicefrac{-B}{2},\nicefrac{B}{2}]$ and $t \in [\nicefrac{-T}{2},\nicefrac{T}{2}]$. We observe that the proposed IOTA pulse shapes are more localized than the Gaussian-sinc filter~\cite{Chockalingam2025_gs}. The IOTA PSWF pulse is also more localized than the IOTA Gaussian filter due to the optimality of the PSWF pulse in localization~\cite{pollak1961_pswf1,pollak1961_pswf2}.

\subsection{\textcolor{black}{I/O Relation Estimation (Sensing) Performance}}
\label{subsec:io_rel}

\textcolor{black}{Fig.~\ref{fig:nmse} plots the normalized mean squared error (NMSE) for estimating the DD effective channel $\mathbf{h}_{\mathrm{eff}}[k,l]$ (sensing) via~\eqref{eq:sys_model12} using a separate pilot frame with pilot signal-to-noise ratio (SNR) equal to the data SNR. The NMSE is defined as $\text{NMSE} = \frac{\sum_{k,l}|\widehat{\mathbf{h}}_{\mathrm{eff}}[k,l] - \mathbf{h}_{\mathrm{eff}}[k,l]|^{2}}{\sum_{k,l}|\mathbf{h}_{\mathrm{eff}}[k,l]|^{2}}$ for an estimate $\widehat{\mathbf{h}}_{\mathrm{eff}}[k,l]$ of the DD effective channel. We observe that maximum localization of the Gaussian and RRC pulses (due to $1.6 \times$ time and bandwidth expansion) results in the best NMSE performance (exponential reduction with SNR), whereas the non-localized behavior of the sinc pulse results in the worst NMSE performance (plateaus at moderate SNRs). The Gaussian-sinc filter~\cite{Chockalingam2025_gs} achieves significantly better NMSE compared to the sinc filter; however, since it is not strictly orthogonal, the NMSE plateaus at high SNRs. On the other hand, the improved localization of the proposed IOTA pulses over the Gaussian-sinc pulse (see Fig.~\ref{fig:pulse}) results in improved NMSE performance close to the Gaussian filter, with exponential reduction vs SNR.} 


\subsection{\textcolor{black}{Data Detection Performance}}
\label{subsec:data_det}

\textcolor{black}{Figs.~\ref{fig:ber}(\subref{fig:ber_perfectcsi})-(\subref{fig:ber_estcsi}) plot the bit error rate (BER) for uncoded $4$-QAM (quadrature amplitude modulation) transmissions for perfect and estimated I/O relation knowledge at the receiver respectively. I/O relation estimation is performed similar to Section~\ref{subsec:io_rel}, i.e., using a separate pilot frame with pilot SNR equal to the data SNR. Note that the Gaussian filter has poor BER due to its non-orthogonality, and the poor NMSE for I/O relation with the sinc filter (Fig.~\ref{fig:nmse}) results in poor BER with estimated I/O relation. Due to their orthogonality, localized behavior, and improved NMSE performance from Fig.~\ref{fig:nmse}, the proposed IOTA pulses offer ideal BER performance with both perfect and estimated I/O relation, with improved performance due to lower NMSE over the Gaussian-sinc filter from~\cite{Chockalingam2025_gs}.}

\subsection{\textcolor{black}{Spectral Efficiency Comparison of Various Pulse Shapes}}
\label{subsec:se}

\begin{figure}
    \centering
    \includegraphics[width=0.95\linewidth]{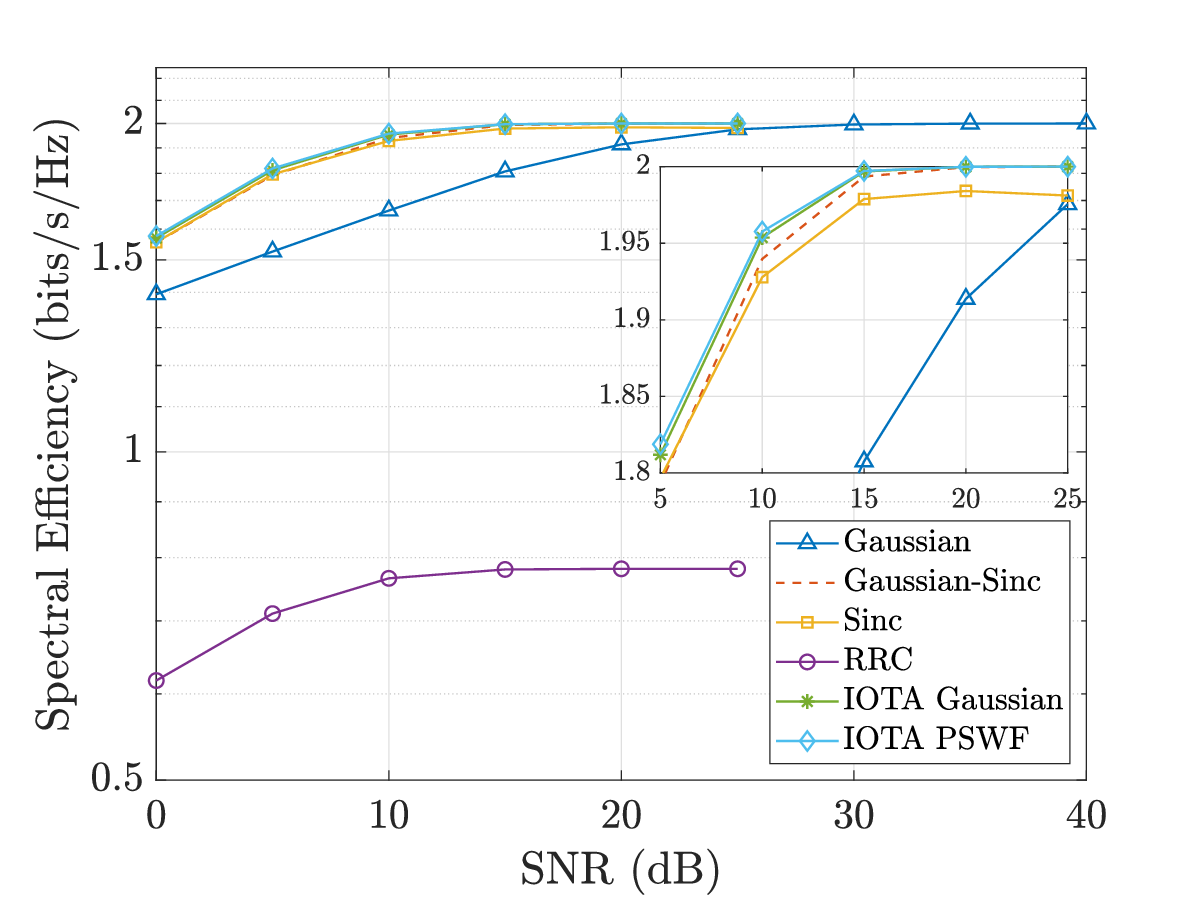}
    \caption{\textcolor{black}{Spectral efficiency of various pulse shaping filters for uncoded $4$-QAM transmission with I/O relation estimation.}}
    \label{fig:se}
\end{figure}

\textcolor{black}{Fig.~\ref{fig:se} compares the spectral efficiency (SE) achieved by different pulse shapes for uncoded $4$-QAM transmission with estimated I/O relation. Due to $1.6 \times$ time and bandwidth expansion, the RRC filter achieves the lowest SE, with a maximum value of $\nicefrac{\log_{2}{4}}{1.6^2} \approx 0.78$ bits/s/Hz. Due to its non-orthogonality and poor BER performance in Fig.~\ref{fig:ber}, the Gaussian filter achieves the maximum possible SE of $2$ bits/s/Hz only at $30$ dB SNR. Due to poor I/O relation estimation and BER saturation at $10^{-2}$ in Fig.~\ref{fig:ber}(\subref{fig:ber_estcsi}), the sinc filter achieves a maximum SE of $0.99 \times 2 = 1.98$ bits/s/Hz. The proposed IOTA filters achieve the best SE performance, with improvements of about $1 \%$ over the Gaussian-sinc SE at $10$ dB SNR.}

\section{Conclusion}
\label{sec:conclusion}

In this paper, we applied the Isotropic Orthogonal Transform Algorithm to the design of delay-Doppler pulse shaping filters for integrated sensing \& communication. Unlike previous approaches, the proposed filters are localized, achieve full spectral efficiency and do not suffer from inter-symbol interference. \textcolor{black}{Future work will investigate the interplay between pulse shaping and coding, and explore extensions to multi-user MIMO and applications such as sidelink-based sensing}.


\bibliographystyle{IEEEtran}
\bibliography{references}

@misc{EURASIP2025,
      title={{Discrete Radar based on Modulo Arithmetic}}, 
      author={Nishant Mehrotra and Sandesh Rao Mattu and Saif Khan Mohammed and Ronny Hadani and Robert Calderbank},
      year={2025},
      eprint={2508.15671},
      archivePrefix={arXiv},
      primaryClass={eess.SP},
      url={https://arxiv.org/abs/2508.15671}, 
}

@ARTICLE{Mehrotra2025_WCLSpread,
  author={Mehrotra, Nishant and Mattu, Sandesh Rao and Calderbank, Robert},
  journal={IEEE Wireless Communications Letters}, 
  title={{Zak-OTFS With Spread Carrier Waveforms}}, 
  year={2025},
  volume={14},
  number={10},
  pages={3244-3248},
  doi={10.1109/LWC.2025.3590254}
}

@book{otfs_book,
  author={Mohammed, Saif Khan and Hadani, Ronny and Chockalingam, Ananthanarayanan},
  title={{OTFS} {M}odulation: {T}heory and {A}pplications}, 
  publisher={Wiley-IEEE Press},
  year={2024},
  volume={},
  number={},
  address = {Hoboken, NJ},
  doi={10.1002/9781119984221}
}

@ARTICLE{bitspaper1,
  author={Mohammed, Saif Khan and Hadani, Ronny and Chockalingam, Ananthanarayanan and Calderbank, Robert},
  journal={IEEE BITS the Information Theory Magazine}, 
  title={{{OTFS}—A Mathematical Foundation for Communication and Radar Sensing in the Delay-{D}oppler Domain}}, 
  year={2022},
  volume={2},
  number={2},
  pages={36-55},
  doi={10.1109/MBITS.2022.3216536}
}

@ARTICLE{bitspaper2,
  author={Mohammed, Saif Khan and Hadani, Ronny and Chockalingam, Ananthanarayanan and Calderbank, Robert},
  journal={IEEE BITS the Information Theory Magazine}, 
  title={{{OTFS}—Predictability in the Delay-{D}oppler Domain and Its Value to Communication and Radar Sensing}}, 
  year={2023},
  volume={3},
  number={2},
  pages={7-31},
  doi={10.1109/MBITS.2023.3319595}
}

@misc{Mohammed2024_pulseshaping,
  title={{Zak-OTFS: Pulse Shaping and the Tradeoff between Time/Bandwidth Expansion and Predictability}}, 
  author={Jinu Jayachandran and Rahul Kumar Jaiswal and Saif Khan Mohammed and Ronny Hadani and Ananthanarayanan Chockalingam and Robert Calderbank},
  year={2024},
  eprint={2405.02718},
  archivePrefix={arXiv},
  primaryClass={eess.SP},
  url={https://arxiv.org/abs/2405.02718}, 
}

@ARTICLE{Calderbank2025_interleaved,
  author={Jayachandran, Jinu and Khan, Imran Ali and Mohammed, Saif Khan and Hadani, Ronny and Chockalingam, Ananthanarayanan and Calderbank, Robert},
  journal={IEEE Transactions on Vehicular Technology}, 
  title={{Zak-OTFS with Interleaved Pilots to Extend the Region of Predictable Operation}}, 
  year={2025},
  volume={},
  number={},
  pages={1-15},
  doi={10.1109/TVT.2025.3579394}
}

@misc{Calderbank2025_isac,
    title={{Zak-OTFS to Integrate Sensing the I/O Relation and Data Communication}}, 
    author={Muhammad Ubadah and Saif Khan Mohammed and Ronny Hadani and Shachar Kons and Ananthanarayanan Chockalingam and Robert Calderbank},
    year={2025},
    eprint={2404.04182},
    archivePrefix={arXiv},
    primaryClass={eess.SP},
    url={https://arxiv.org/abs/2404.04182}, 
}

@misc{zakotfs_ltv,
      title={{Zak-OTFS for Identification of Linear Time-Varying Systems}}, 
      author={Danish Nisar and Saif Khan Mohammed and Ronny Hadani and Ananthanarayanan Chockalingam and Robert Calderbank},
      year={2025},
      eprint={2503.18900},
      archivePrefix={arXiv},
      primaryClass={eess.SP},
      url={https://arxiv.org/abs/2503.18900}, 
}

@ARTICLE{Chockalingam2025_gs,
  author={Das, Arpan and Jesbin, Fathima and Chockalingam, Ananthanarayanan},
  journal={IEEE Transactions on Vehicular Technology}, 
  title={{A Gaussian-Sinc Pulse Shaping Filter for Zak-OTFS}}, 
  year={2025},
  volume={},
  number={},
  pages={1-16},
  doi={10.1109/TVT.2025.3609781}
}

@misc{Chockalingam2025_hermite,
  title={{Delay-Doppler Pulse Shaping in Zak-OTFS Using Hermite Basis Functions}}, 
  author={Fathima Jesbin and Ananthanarayanan Chockalingam},
  year={2025},
  eprint={2510.17466},
  archivePrefix={arXiv},
  primaryClass={cs.IT},
  url={https://arxiv.org/abs/2510.17466}, 
}

@ARTICLE{Gopalam2024_tfwindowing,
  author={Gopalam, Swaroop and Collings, Iain B. and Hanly, Stephen V. and Inaltekin, Hazer and Pillai, Sibi Raj B. and Whiting, Philip},
  journal={IEEE Transactions on Communications}, 
  title={{Zak-OTFS Implementation via Time and Frequency Windowing}}, 
  year={2024},
  volume={72},
  number={7},
  pages={3873-3889},
  doi={10.1109/TCOMM.2024.3366403}
}

@misc{Calderbank2025_uplinkrac,
  title={{Zak-OTFS Based Coded Random Access for Uplink mMTC}}, 
  author={Alessandro Mirri and Venkatesh Khammammetti and Beyza Dabak and Enrico Paolini and Krishna Narayanan and Robert Calderbank},
  year={2025},
  eprint={2507.22013},
  archivePrefix={arXiv},
  primaryClass={eess.SP},
  url={https://arxiv.org/abs/2507.22013}, 
}

@ARTICLE{Berrou1995_iota,
  author={Le Floch, B. and Alard, M. and Berrou, C.},
  journal={Proceedings of the IEEE}, 
  title={{Coded Orthogonal Frequency Division Multiplex}}, 
  year={1995},
  volume={83},
  number={6},
  pages={982-996},
  doi={10.1109/5.387096}
}

@ARTICLE{Strohmer2003_iota,
  author={Strohmer, T. and Beaver, S.},
  journal={IEEE Transactions on Communications}, 
  title={{Optimal OFDM Design for Time-Frequency Dispersive Channels}}, 
  year={2003},
  volume={51},
  number={7},
  pages={1111-1122},
  doi={10.1109/TCOMM.2003.814200}
}

@article{Belfiore1997_hermite,
  title={{A Time-Frequency Well-Localized Pulse for Multiple Carrier Transmission}},
  author={Haas, Ralf and Belfiore, Jean-Claude},
  journal={Wireless personal communications},
  volume={5},
  number={1},
  pages={1--18},
  year={1997},
  publisher={Springer}
}

@ARTICLE{Farhang2015_fbmc,
  author={Amini, Pooyan and Chen, Rong-Rong and Farhang-Boroujeny, Behrouz},
  journal={IEEE Journal of Oceanic Engineering}, 
  title={{Filterbank Multicarrier Communications for Underwater Acoustic Channels}}, 
  year={2015},
  volume={40},
  number={1},
  pages={115-130},
  doi={10.1109/JOE.2013.2291139}
}

@ARTICLE{Arslan2014_fbmcsurvey,
  author={Sahin, Alphan and Guvenc, Ismail and Arslan, Huseyin},
  journal={IEEE Communications Surveys \& Tutorials}, 
  title={{A Survey on Multicarrier Communications: Prototype Filters, Lattice Structures, and Implementation Aspects}}, 
  year={2014},
  volume={16},
  number={3},
  pages={1312-1338},
  doi={10.1109/SURV.2013.121213.00263}
}

@ARTICLE{pollak1961_pswf1,
  author={Slepian, D. and Pollak, H. O.},
  journal={The Bell System Technical Journal}, 
  title={{Prolate Spheroidal Wave Functions, Fourier Analysis and Uncertainty — I}}, 
  year={1961},
  volume={40},
  number={1},
  pages={43-63},
  doi={10.1002/j.1538-7305.1961.tb03976.x}
}

@ARTICLE{pollak1961_pswf2,
  author={Landau, H. J. and Pollak, H. O.},
  journal={The Bell System Technical Journal}, 
  title={{Prolate Spheroidal Wave Functions, Fourier Analysis and Uncertainty — II}}, 
  year={1961},
  volume={40},
  number={1},
  pages={65-84},
  doi={10.1002/j.1538-7305.1961.tb03977.x}
}

@article{veh_a,
    author={{ITU--R M}.1225},
    year = {1997},
    title = {{Guidelines for evaluation of radio transmission technologies for IMT-2000}},
    journal = {International Telecommunication Union Radio communication}
}

@ARTICLE{Jamalipour2024,
  author={Liu, Ruiqi and Hua, Meng and Guan, Ke and Wang, Xiping and Zhang, Leyi and Mao, Tianqi and Zhang, Di and Wu, Qingqing and Jamalipour, Abbas},
  journal={IEEE Transactions on Intelligent Transportation Systems}, 
  title={{6G Enabled Advanced Transportation Systems}}, 
  year={2024},
  volume={25},
  number={9},
  pages={10564-10580},
  doi={10.1109/TITS.2024.3362515}
}
\end{document}